%% file: main.tex
\documentclass[sigplan,screen,nonacm]{acmart}

\AtBeginDocument{%
  \providecommand\BibTeX{{%
    \normalfont B\kern-0.5em{\scshape i\kern-0.25em b}\kern-0.8em\TeX}}}

\setcopyright{none}
\acmConference[Preprint]{Preprint}{}{}
\acmBooktitle{Preprint}
\usepackage{subcaption}
\usepackage{booktabs}
\usepackage{tabularx}

\makeatletter
\newcommand\footnoteref[1]{\protected@xdef\@thefnmark{\ref{#1}}\@footnotemark}
\makeatother

\usepackage{microtype}
\usepackage{listings}
\usepackage{tikz}
\usepackage{pifont}
\usetikzlibrary{shapes}

\DeclareRobustCommand*\circledColorSmall[2]{\tikz[baseline=(char.base)]{
    \node[shape=circle,fill=#2,draw=#2,inner sep=0pt] (char) {\textcolor{white}{\footnotesize\textbf{#1}}};}}
\lstdefinestyle{C}{
    language=C,
    keywordstyle=\color{blue},
    stringstyle=\color{red},
    commentstyle=\color{olive},
}

\lstdefinestyle{tinyC}{
    language=C,
    basicstyle={\tiny\ttfamily},
    keywordstyle=\color{blue},
    stringstyle=\color{red},
    commentstyle=\color{olive},
}

\lstdefinestyle{tinyPy}{
    language=Python,
    basicstyle={\tiny\ttfamily},
    keywordstyle=\color{blue},
    stringstyle=\color{red},
    commentstyle=\color{olive},
}

\lstdefinestyle{JSON}{
    sensitive=false,
    basicstyle={\tiny\ttfamily},
    stringstyle=\color{red},
    morestring=[b]"
}

\lstdefinestyle{MLIR}{
    sensitive=false,
    stringstyle=\color{red},
    string=[b]",
    commentstyle=\color{olive},
    comment=[l]{//},
    keywordstyle=[1]\color{blue},
    keywordstyle=[2]\color{violet},
    keywordstyle=[3]\color{brown},
    alsoletter={\%},
    keywords=[1]{sdfg, arith, memref, func, return},
    keywords=[2]{
        \%a, \%b, \%c, \%d, \%i, \%r,
        \@state_0, \@state_1, \@s0, \@s1, \@s2, \@s3, \@fName,
        \@init_2, \@constant_3, \@load_6, \@load_8, \@addi_9, \@return_11,
        \%arg0, \%arg1, \%arg2, \%arg3, \%arg4, \%arg5, \%arg6, \%arg7, \%arg8, \%arg9, \%arg10,
        \%c0, 
        \%0, \%1, \%2, \%3, \%4, \%5, \%6, \%7, \%8, \%9, \%10
    },
    keywords=[3]{i32, xi32, index},
}

\lstdefinestyle{tinyMLIR}{
    sensitive=false,
    basicstyle={\tiny\ttfamily},
    stringstyle=\color{red},
    string=[b]",
    commentstyle=\color{olive},
    comment=[l]{//},
    keywordstyle=[1]\color{blue},
    keywordstyle=[2]\color{violet},
    keywordstyle=[3]\color{brown},
    alsoletter={\%},
    escapechar=\$,
    keywords=[1]{sdfg, arith, memref, func, return},
    keywords=[2]{
        \%a, \%b, \%c, \%d, \%i, \%r,
        \@state_0, \@state_1, \@s0, \@s1, \@s2, \@s3, \@fName,
        \@init_2, \@constant_3, \@load_6, \@load_8, \@addi_9, \@return_11,
        \%arg0, \%arg1, \%arg2, \%arg3, \%arg4, \%arg5, \%arg6, \%arg7, \%arg8, \%arg9, \%arg10,
        \%c0, 
        \%0, \%1, \%2, \%3, \%4, \%5, \%6, \%7, \%8, \%9, \%10
    },
    keywords=[3]{i32, xi32, index},
}

\lstdefinestyle{MLIR-Table}{
    basicstyle={\footnotesize\ttfamily},
    sensitive=false,
    stringstyle=\color{red},
    string=[b]",
    commentstyle=\color{olive},
    comment=[l]{//},
    keepspaces=true,
    keywordstyle=[1]\color{blue},
    keywordstyle=[2]\color{violet},
    keywordstyle=[3]\color{brown},
    alsoletter={\%},
    keywords=[1]{sdfg, arith},
    keywords=[2]{
        \%a, \%b, \%c, \%d, \%i, \%r,
        \@state_0, \@state_1, \@s0, \@s1, \@s2, \@s3,
    },
    keywords=[3]{i32, xi32, index},
}

\newcommand{\macsection}[1]{\textbf{\textit{#1.}}~~}
    
\begin{document}

\title{Bridging Control-Centric and Data-Centric Optimization}

\author{Tal Ben-Nun}
\authornote{Both authors contributed equally to the paper.}
\affiliation{%
  \institution{ETH Zurich}
  \city{Zurich}
  \country{Switzerland}
}
\email{talbn@inf.ethz.ch}

\author{Berke Ates}
\authornotemark[1]
\affiliation{%
  \institution{ETH Zurich}
  \city{Zurich}
  \country{Switzerland}
}
\email{beates@student.ethz.ch}

\author{Alexandru Calotoiu}
\affiliation{%
  \institution{ETH Zurich}
  \city{Zurich}
  \country{Switzerland}
}
\email{acalotoiu@inf.ethz.ch}

\author{Torsten Hoefler}
\affiliation{%
  \institution{ETH Zurich}
  \city{Zurich}
  \country{Switzerland}
}
\email{htor@inf.ethz.ch}

\input{sections/abstract}

\input{sections/introduction}
\input{sections/background}

\input{sections/data-centric_dialect}
\input{sections/transformations}

\input{sections/evaluation}

\input{sections/related_work}
\input{sections/conclusion}

\input{sections/acknowledgements}
\input{sections/artifact.tex}

\bibliographystyle{ACM-Reference-Format}
\bibliography{refs}

\end{document}

%% file: sections/abstract.tex
\begin{abstract}
With the rise of specialized hardware and new programming languages, code optimization has shifted its focus towards promoting data locality.
Most production-grade compilers adopt a control-centric mindset --- instruction-driven optimization augmented with scalar-based dataflow --- whereas other approaches provide domain-specific and general purpose data movement minimization, which can miss important control-flow optimizations.
As the two representations are not commutable, users must choose one over the other.
In this paper, we explore how both control- and data-centric approaches can work in tandem via the Multi-Level Intermediate Representation (MLIR) framework.
Through a combination of an MLIR dialect and specialized passes, we recover parametric, symbolic dataflow that can be optimized within the DaCe framework.
We combine the two views into a single pipeline, called DCIR, showing that it is strictly more powerful than either view.
On several benchmarks and a real-world application in C, we show that our proposed pipeline consistently outperforms MLIR and automatically uncovers new optimization opportunities with no additional effort.
\end{abstract}

\ccsdesc[500]{Software and its engineering~Correctness}
\ccsdesc[500]{Software and its engineering~Software performance}
\ccsdesc[500]{Software and its engineering~Compilers}

\keywords{
MLIR, DaCe, data-centric programming
}

\maketitle

%% file: sections/introduction.tex
\begin{figure}[h]
    \centering
    \vspace{-1em}
    \includegraphics[width=.85\linewidth]{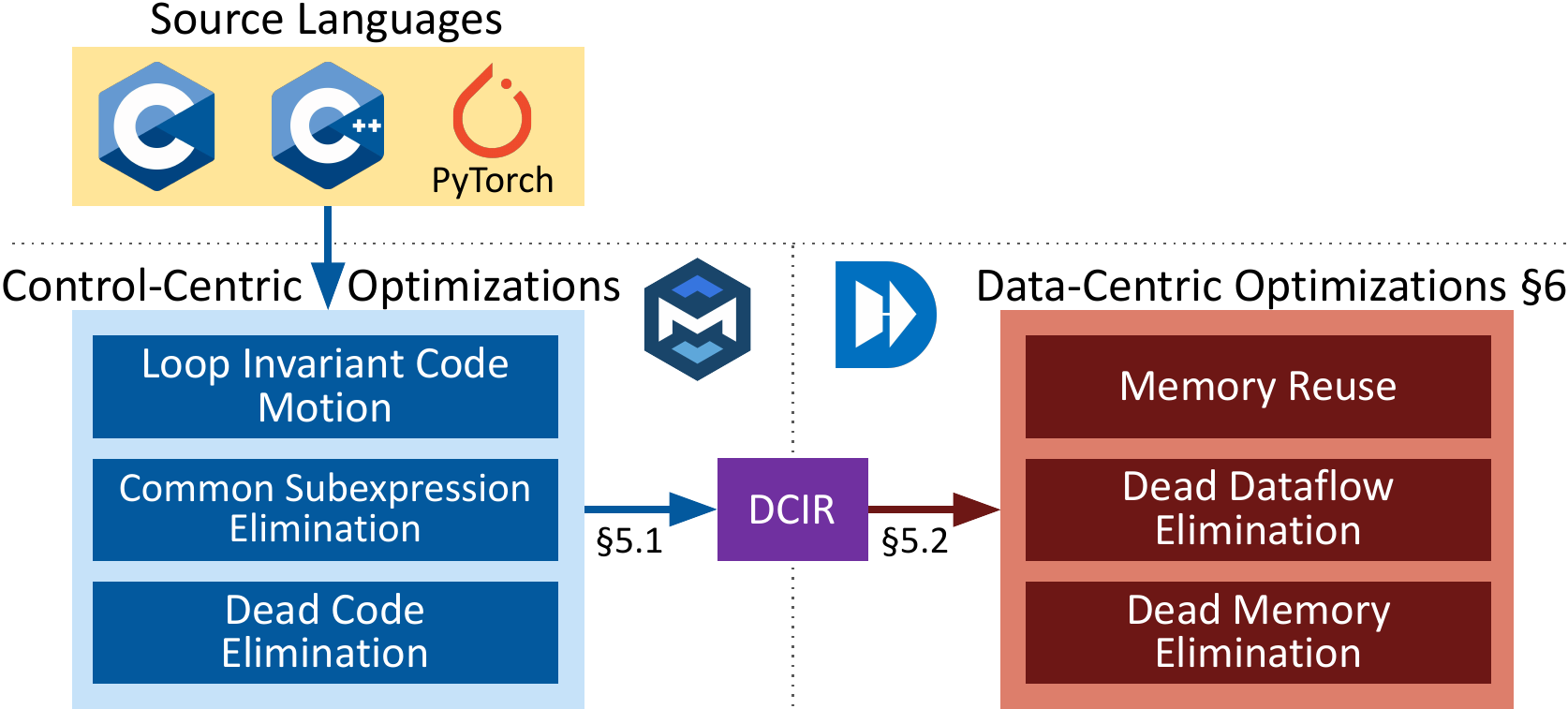}
    \vspace{-1em}
    \caption{Overview of DCIR.}
    \vspace{-1.5em}
    \label{fig:introduction_overview}
\end{figure}

\section{Introduction}
Overcoming general-purpose optimization barriers is supported today by software optimization frameworks and intermediate representations (IRs).
Whether driven by domain-specific optimizations~\cite{tc,tvm,taco,oec} or by specializing to emergent hardware~\cite{llhd,circt}, each such framework delivers benefits that cannot be covered by general compilers. 

The Multi-Level Intermediate Representation (MLIR)~\cite{mlir} project aims to standardize such efforts via the use of \textit{dialects} and a homogenized pass infrastructure.
Some classes of full-program optimizations, however, cannot natively fall into the category of a dialect. Data locality within operators, but also across entire applications, becomes crucial to manage~\cite{padal}, which led to the emergence of \textit{data-centric} abstractions~\cite{dace,gunrock,legion,bamboo} and optimization methodologies~\cite{dmiayn}. Such abstractions are promising in that they can verify programs in novel ways (bounds analysis, potential race condition detection), and mutate data movement (global layout management, distributed domain decomposition, local caching, etc.) and allocation (e.g., eliding memory subregions or entire arrays). However, to operate, data-centric optimizations require the visibility, whether constant or parametric, of all memory accesses for their analysis.

In this work, we build a general-purpose bridging infrastructure between existing compiler infrastructure and data-centric representations, called \textbf{DCIR}, benefiting from both aspects in a single compilation pipeline (Fig.~\ref{fig:introduction_overview}). MLIR serves as a natural ``connective tissue'', with lowering and conversion from frontends in different languages and to backends covering a plethora of hardware architectures. For the data-centric representation, we choose the DaCe framework and its stateful dataflow multigraph (SDFG) IR~\cite{dace}, which is capable of optimizing a wide variety of applications through data movement minimization~\cite{omen,fv3,soap,daceml,p3dace,c2dace,stencilflow}.

\begin{figure}[t]
    \centering
    \begin{subfigure}{.8\linewidth}
\begin{lstlisting}[style=tinyC]
int example() {
  int *A = (int*)malloc(100000 * sizeof(int));
  int *B = (int*)malloc(100000 * sizeof(int));
  for (int i = 0; i < 100000; ++i) {
    A[i] = 5;
    for (int j = 0; j < 100000; ++j)
      B[j] = A[i];
    for (int j = 0; j < 10000; ++j) 
      A[j] = A[i];
  }
  int res = B[0];
  free(A);
  free(B);
  return res;
}
\end{lstlisting}
        \vspace{-0.5em}
        \caption{Input code}
    \end{subfigure}
    \begin{subfigure}{\linewidth}
        \centering
        \includegraphics[width=.8\textwidth]{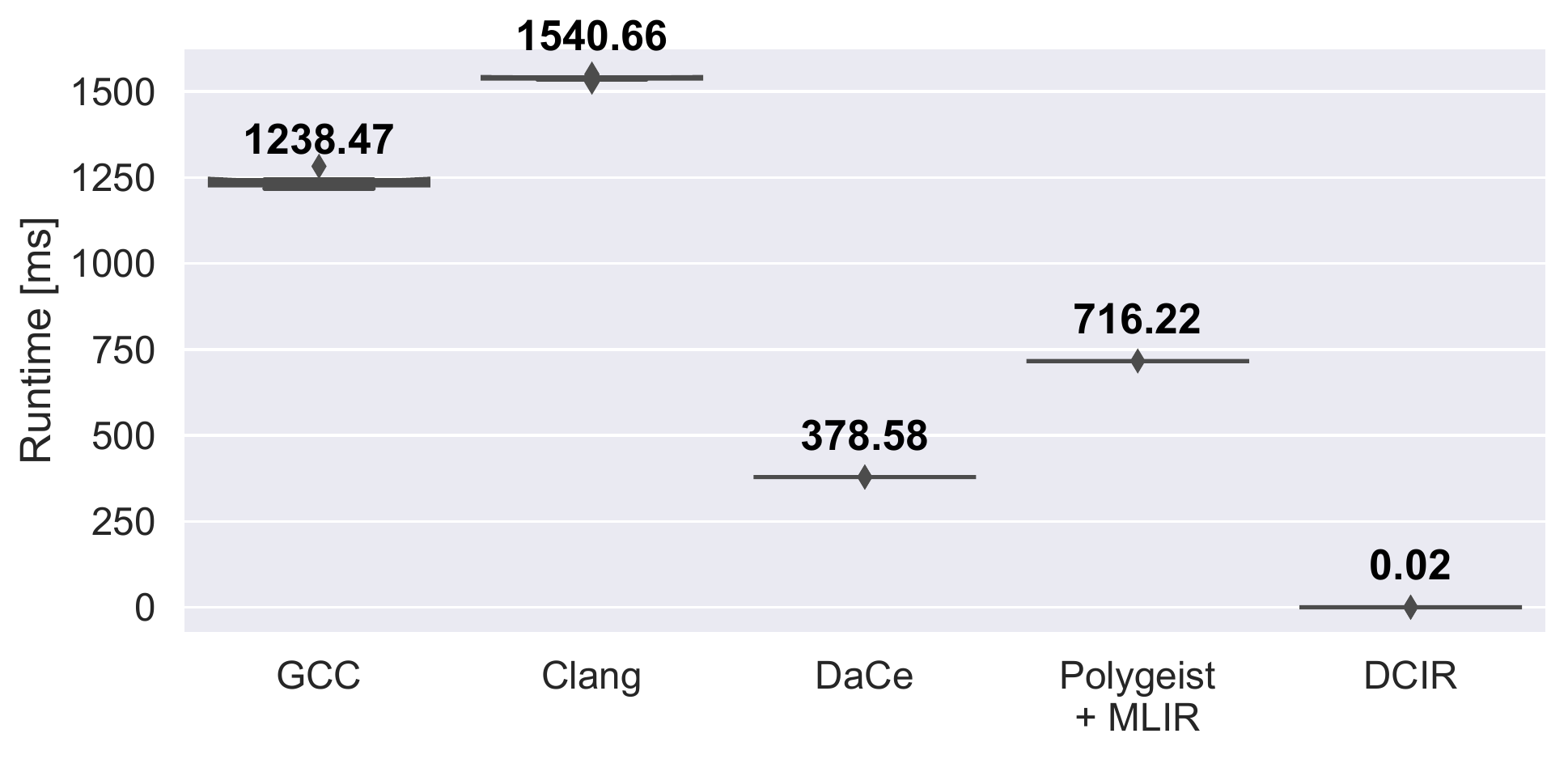}
        \vspace{-0.5em}
        \caption{Performance across compilers}
    \end{subfigure}
    \vspace{-1em}
    \caption{Mixed control- and data-centric analysis.}
    \vspace{-1em}
    \label{fig:introduction_motivation}
\end{figure}

The usefulness of mixed-mode optimization can be demonstrated with a simple C example, shown in Fig.~\ref{fig:introduction_motivation}.
Both production compilers and data-centric IRs produce suboptimal code on their own. Optimizations applied by GCC and LLVM are able to fuse the first two loops, setting every entry of \texttt{A} and \texttt{B} to 5, with MLIR further optimizing the write schedule to \texttt{A}. However, the third loop remains intact, although the entire memory allocation is unnecessary. Similarly, the DaCe framework can potentially elide the third loop, but requires global code motion to find the false dependency across the two arrays.
By combining both optimization classes, DCIR elides all loops, reducing the code to a single statement.

To convert existing MLIR dialects to the SDFG IR, we augment global data movement analysis in MLIR. DCIR employs a multi-stage approach on both IRs, starting from introducing symbolic sizes to MLIR \texttt{memref}s and propagating them to recover parametric subsets of moved data. We then lift and regroup computations to mitigate reliance on link-time optimization. Lastly, we introduce specialized data-centric transformations to the DaCe framework to better expose statically analyzable data movement generated by MLIR.

To evaluate the representational strength of mixed-mode optimization, we run a variety of applications with a standard optimizing pipeline (\texttt{-O2}, MLIR$\rightarrow$SDFG) and \textit{no domain-specific techniques}. The results indicate that the DCIR pipeline can recover the semantics necessary for standard MLIR dialects to match and outperform the original input codes, as well as expose new optimization opportunities automatically. The paper makes the following contributions:
\begin{itemize}
    \item Introducing data-centric optimizations to a standard multi-level compilation pipeline through an MLIR dialect (\S\ref{sec:dialect}) and a complementary set of passes (\S\ref{sec:passes});
    \item Addressing MLIR limitations that inhibit global data movement analysis (\S\ref{sec:symbols}) and efficiency (\S\ref{sec:polybench});
    \item Demonstrating performance results (\S\ref{sec:evaluation}) on small and real-world benchmarks, exhibiting 1.59$\times$ geomean\\ speedup in Polybench/C over MLIR, 2.33$\times$ speedup for a deep learning activation function, and 7$\times$ speedup in MILC over the best general-purpose compiler.
\end{itemize}

%% file: sections/background.tex
\section{Background} 
We first discuss the approaches we build upon, namely the MLIR~\cite{mlir} and DaCe~\cite{dace} frameworks. 

\subsection{MLIR}

Most compiler frameworks such as LLVM~\cite{llvm} and JVM~\cite{jvm} rely on a single abstraction level --- programs are translated from a given language into an intermediate representation, and all transformation, optimization, and verification happens on it. By providing a framework in which multiple level of abstraction can coexist, MLIR~\cite{mlir} aims to make the development and interoperability of different IRs easier.

An intermediate representation within MLIR is called a {\em dialect}. Dialects provide specialized operators and data types, as well as conversion points from higher-level IRs (lowering) and equivalent-level IRs (converters). MLIR contains several core dialects and the rest are defined by users. Most dialects provide lowering capabilities one level down, thus saving redundant work in compiler infrastructure development.

MLIR is gaining community traction on front and back ends. Frontend projects such as Flang~\cite{flang} translate FORTRAN code to MLIR dialects. For backends, Katel et al.~\cite{mlir-gpu} recently conducted a performance evaluation comparing MLIR to CUBLAS~\cite{cublas} on matrix-matrix multiplications and showed that MLIR could generate competitive code for GPUs.

In our workflow, we leverage Polygeist~\cite{polygeistPACT}, a prototype frontend for MLIR that is capable of translating subsets of C and C++ to MLIR dialects.
Polygeist emits MLIR code using the structured control flow (\texttt{scf}), arithmetic (\texttt{arith}), and the memory reference (\texttt{memref}) dialects.
These three dialects are sufficient for Polygeist to express most C constructs: the structured control flow dialect encompasses loops and branch constructs, the arithmetic dialect can be used to express arithmetic and logical operations, and the \texttt{memref} dialect describes data layout, array dimensions, and sizes.
While Polygeist is not currently sufficiently stable to translate large scientific benchmarks, it can handle the Polybench suite~\cite{polybench} and snippets of real-world programs (\S\ref{sec:evaluation}).

\subsection{DaCe}
\label{sec:dace}

There are many optimization techniques aimed at improving data movement, though most are specific to particular hardware systems or even individual hardware configurations. At the same time, requiring high-level programs to consider data movement comes at the cost of added code complexity. 

DaCe is a programming framework that addresses this issue by defining an IR called Stateful DataFlow multiGraphs (SDFG)~\cite{dace}, built around understanding data movement. Da-Ce provides frontends to translate code written in Python, Octave, or C into the SDFG IR. It also provides a transformation API on the IR to separate the concerns of the developer and the performance engineer. 
DaCe has succesfully improved the performance of applications in weather and climate models~\cite{stencilflow,fv3}, sparse linear algebra in quantum transport simulation~\cite{omen}, graph analytics~\cite{dace}, and full neural network optimization in deep learning~\cite{daceml}. 

Performance benefits in DaCe stem from a combination of local and global optimizations. Local optimizations are expressed as graph rewriting transformations, which perform a variety of operations: adding or removing explicit memory allocation, e.g., for increased cache utilization; SIMD vectorization; tiling parallel sections; and converting random-access memory into streaming (FIFO queues) when beneficial. For global optimizations, the same techniques exposed by the SDFG IR enable changing communication schemes in distributed applications and array layouts based on data movement modeling, or memory footprint reduction via memory region ``live-set'' analysis.

Optimization in the DaCe framework start by an SDFG simplification pass, which enlarges the pure dataflow regions and removes redundant memory allocation, equivalent to \texttt{-O1} in compilers. Subsequently, optimizations are applied via automated heuristics (\texttt{-O2}), followed by potential manual application of transformations by performance engineers.

The SDFG IR itself is represented by a control-flow graph (state machine) of dataflow acyclic multigraphs. SDFGs separate data containers and data movement (represented as dataflow edges called \textit{memlets}) from their use in computational nodes (called \textit{tasklets}). This allows the IR to express data movement explicitly --- and even relies on data dependencies to create the execution order. Explicit control-flow is only used when dataflow cannot be otherwise inferred. 

An important tool the SDFG IR relies on is parametric representation of data access patterns. By using symbolic expressions to represent array accesses, it becomes possible, for example, to determine whether or not accesses may overlap (whether sparse/indirect or dense).

Conversely, SDFGs cannot natively represent passes such as loop-invariant code motion or general common subexpression elimination, which control-centric IRs routinely implement. Because the tasklet is seen as an atomic unit, its contents cannot be inspected for transformations.
Additionally, some design patterns cannot be represented concisely in SDFGs, such as parametric-depth recursion or dynamic pointers, which could be adequately handled by other IRs. Therefore, a bridge between SDFG and MLIR is a logical choice that can aid optimization on both ends.

%% file: sections/data-centric_dialect.tex
\section{Data-Centric MLIR Dialect} \label{sec:dialect}

At the core of our proposed bridge --- DCIR --- lies the \texttt{sdfg} dialect. Its purpose is to exist as a convertible target to/from the standard MLIR dialects within the MLIR dialect conversion framework, as well as a representation that is directly translatable to the DaCe SDFG IR. Conversion to the \texttt{sdfg} dialect, however, is not straightforward due to fundamental representational differences:
\begin{enumerate}
    \item All SDFG data containers must be defined with a predetermined constant or parametric size;
    \item Data movement granularity (subregions, indices) must be specified at each scope;
    \item Memory operations in SDFG are divided into load, store, and update, as opposed to the load/store view in standard MLIR dialects.
\end{enumerate}

\begin{figure}[t]
    \centering
    \begin{subfigure}{\linewidth}
\begin{lstlisting}[style=MLIR]
func @fName (%A: memref<?xi32>, 
             %B: memref<?xi32>) {
 memref.copy %A, %B : memref<?xi32>
 return
}
\end{lstlisting}
        \vspace{-0.5em}
        \caption{Function that copies the entries of one \texttt{memref} to another}
        \label{fig:mlir-dace_size_validation_memref}
    \end{subfigure}
    \vskip\baselineskip
    \begin{subfigure}{\linewidth}
\begin{lstlisting}[style=MLIR]
func @fName (%A: sdfg.array<sym("2*N")xi32>, 
             %B: sdfg.array<sym("N")xi32>) {
 sdfg.copy %A, %B : !sdfg.array<sym("N")xi32>
 return
}
\end{lstlisting}
        \vspace{-0.5em}
        \caption{Symbolic version of the same function detects the size mismatch}
        \label{fig:mlir-dace_size_validation_sdfg}
    \end{subfigure}
    \vspace{-1em}
    \caption{Parametric size verification with the \texttt{sdfg} dialect.}
    \label{fig:mlir-dace_size_validation}
\end{figure}
\begin{table*}[t]
  \centering
  \footnotesize
  \caption{\texttt{sdfg} Dialect Operations}
  \vspace{-1em}
    \begin{tabularx}{\linewidth}{@{} *5{>{\raggedright\arraybackslash}X} @{}}
        \toprule
            \textbf{Operation} & \textbf{Description} \\
        \midrule
            \lstinline[style=MLIR-Table]!sdfg.tasklet(\%a: i32) -> i32 \{...\}! & \textbf{Computation:} Encapsulated unit of computation, with no external dataflow except for parameters and return values. \\
            
            \addlinespace[0.5em]
            \lstinline[style=MLIR-Table]!sdfg.load \%A[0] : \!sdfg.array<2xi32> -> i32! & \textbf{Loading:} Loads a value from an array. \\
            
            \addlinespace[0.5em]
            \lstinline[style=MLIR-Table]!sdfg.store \%a, \%A[0]: i32 -> \!sdfg.array<2xi32>! & \textbf{Storing and Updating:} Stores a value to an array, or updates it if an update function is given. \\
            
            \addlinespace[0.5em]
            \lstinline[style=MLIR-Table]!sdfg.alloc() : \!sdfg.array<2xi32>! & \textbf{Allocation:} Allocates a data container (array, stream, etc.) of the specified size. May contain symbolic sizes as well. \\
            
            \addlinespace[0.5em]
            \lstinline[style=MLIR-Table]!sdfg.map (\%i) = (0) to (sym("N")) step (1) \ \{...\}! &
            \textbf{Parametric Parallelism:} Represents a scope that is executed in parallel. \\
            
            \addlinespace[0.5em]
            \lstinline[style=MLIR-Table]!sdfg.state @state_0 \{...\}! &
            \textbf{States:} Groups multiple operations. The state machine ensures a correct order of execution and prevents data races.  \\
            
            \addlinespace[0.5em]
            \lstinline[style=MLIR-Table]!sdfg.edge @state_0 -> @state_1! &
            \textbf{State Transition:} Describes the edges of the state machine, linking states together to a directed graph.  \\
        \bottomrule
    \end{tabularx}
  \label{table:mlir-dace-operation-overview}
  \vspace{-1em}
\end{table*}

Since SDFGs rely on parametric dataflow for their analyses (as indicated by the first two differences), we must first augment MLIR to enable expressing both parametric data movement and data container definition.

\subsection{Symbols} \label{sec:symbols}

MLIR introduces the concept of \texttt{memref}s, which enable keeping track of the dimensionality of allocated arrays through multiple levels and enables memory-based optimizations. The memref type allows undefined dimensions with the question mark (\texttt{?}), which is useful to define functions that can work on arrays of arbitrary size. Such a function is presented in Fig.~\ref{fig:mlir-dace_size_validation_memref}, which copies one array's contents to another. However, this question mark also prevents statically checking for any size mismatches in the copy operation. The same issue arises with the star (\texttt{*}) representing arbitrary dimensionality.

DCIR introduces symbolic expressions directly in data type sizes for parametric dataflow analysis.
\textit{Regardless of the dialect}, MLIR inherently disallows defining an identifier used within the types of the function parameters (for example, a function parameter \texttt{\%N} and another parameter of type \texttt{memref<(\%N)xf32>}). We thus resort to maintaining a symbol store and introduce the \texttt{sym} keyword. Symbols are defined globally per module or scope by their name, retaining their (albeit unknown) value throughout their lifetime. Since MLIR disallows defining arbitrary parentheses-enclosed syntax such as \verb|$(N+1)|, nor non-affine expressions, we rely on strings to represent the expressions. Certain scopes (e.g., functions) may accept symbol mappings as attributes to keep correspondence across an application.

Symbolic expressions allow data-centric dialects to perform validation and track unknown sizes during passes. Fig.~\ref{fig:mlir-dace_size_validation} shows a direct comparison of the same function in the \texttt{memref} and \texttt{sdfg} dialects.
Because the \texttt{sdfg.array} type supports symbolic sizes, we can encode the relationship of array sizes while maintaining the flexibility of compile-time undefined sizes. As we can see in Fig.~\ref{fig:mlir-dace_size_validation_sdfg}, the \texttt{sdfg} dialect can statically check for size mismatches and raise an exception at compile-time. This enables catching errors early and avoiding out of bounds access, as well as let passes such as loop tiling track symbolic sizes to produce faster code.

\subsection{Dialect Elements}

We introduce the \texttt{sdfg} dialect, which closely follows the SDFG IR structure, is convertible from the standard dialects, and addresses all three aforementioned representational differences. Its operations are listed in Table~\ref{table:mlir-dace-operation-overview}.

As dictated by the definition of the SDFG IR (\S\ref{sec:dace}), computations (tasklet graph nodes) are defined separately from data movement (edges).
Computations are encapsulated in the \texttt{sdfg.tasklet} operator, which contains an attached scope.
It is an MLIR \texttt{IsolatedFromAbove} scope, which can only operate on scalars and memory references given to it as input arguments. A tasklet can also have zero or more scalars and/or \texttt{memref}s as outputs.
The contents of a tasklet scope can be arbitrary (including loops and other structured control flow constructs), and use any lowerable MLIR dialect, but cannot access any SSA value or memory address outside the ones given to it via memlets. This ensures convertibility to SDFG and analyzability in DaCe. As symbols are read-only throughout their lifetime, they can be readily accessed.

Data movement is abstracted differently than in the SDFG IR. In order to increase readability, as well as maintain a linear-time conversion pass from \texttt{memref}, the dialect provides the \texttt{sdfg.load} and \texttt{sdfg.store} operators. The latter operator can also perform updates (e.g., atomic operations or one-sided communication) via the \texttt{wcr} function attribute (Write-Conflict Resolution in SDFG jargon).

Memory allocation is simplified in our proposed dialect. Since DaCe defines its own allocation lifetime policies (e.g., to save memory footprint), allocation in the generated code is implicit. Thus, all that is necessary in the dialect is to define an \texttt{sdfg.alloc} operation for each data container that DaCe supports. The data container type (e.g., \texttt{sdfg.array}, \texttt{sdfg.stream} for FIFO queues) uses either constants or symbolic expressions to define its size.
Attributes specify the allocation lifetime and whether the allocation management should be performed by the SDFG (called \textit{transient} containers) or not, as well as any aliasing information on the latter.

Parametric parallelism in SDFGs is defined by \textit{map} and \textit{producer/consumer} scopes. The closest equivalent of \texttt{sdfg.map} is \texttt{affine.parallel}, both of which define parallel execution (implemented, e.g., by a GPU kernel, or by FPGA processing elements). Although there is no direct conversion for consume scopes, the construct exists for full commutability between data-centric and control-centric optimizations.

Lastly, when dataflow cannot imply the program order, control-flow constructs are expressed as a finite state machine (similarly to MLIR's \texttt{fsm} dialect). Provided by the \texttt{sdfg.state} and \texttt{sdfg.edge}, structured constructs and general CFGs can be represented via the induced state machine. As opposed to LLVM-style CFGs, edges encode conditions and assignments as symbolic expressions, enabling constant-time testing of data-dependent control flow and retaining semantics of, e.g., switch-case constructs.

In sum, the proposed dialect is, on the one hand, designed for simple conversion from matching components in existing MLIR dialects, and on the other, provides one-to-one compatibility with the SDFG IR. To use the dialect, however, we would need to lift parametric data movement semantics (e.g., symbolic sizes) from existing MLIR codes. Below, we discuss the conversion and translation to the data-centric IR.

\section{Conversion Pipeline}\label{pipeline}

\begin{figure}[t]
    \centering
    \includegraphics[width=0.725\linewidth]{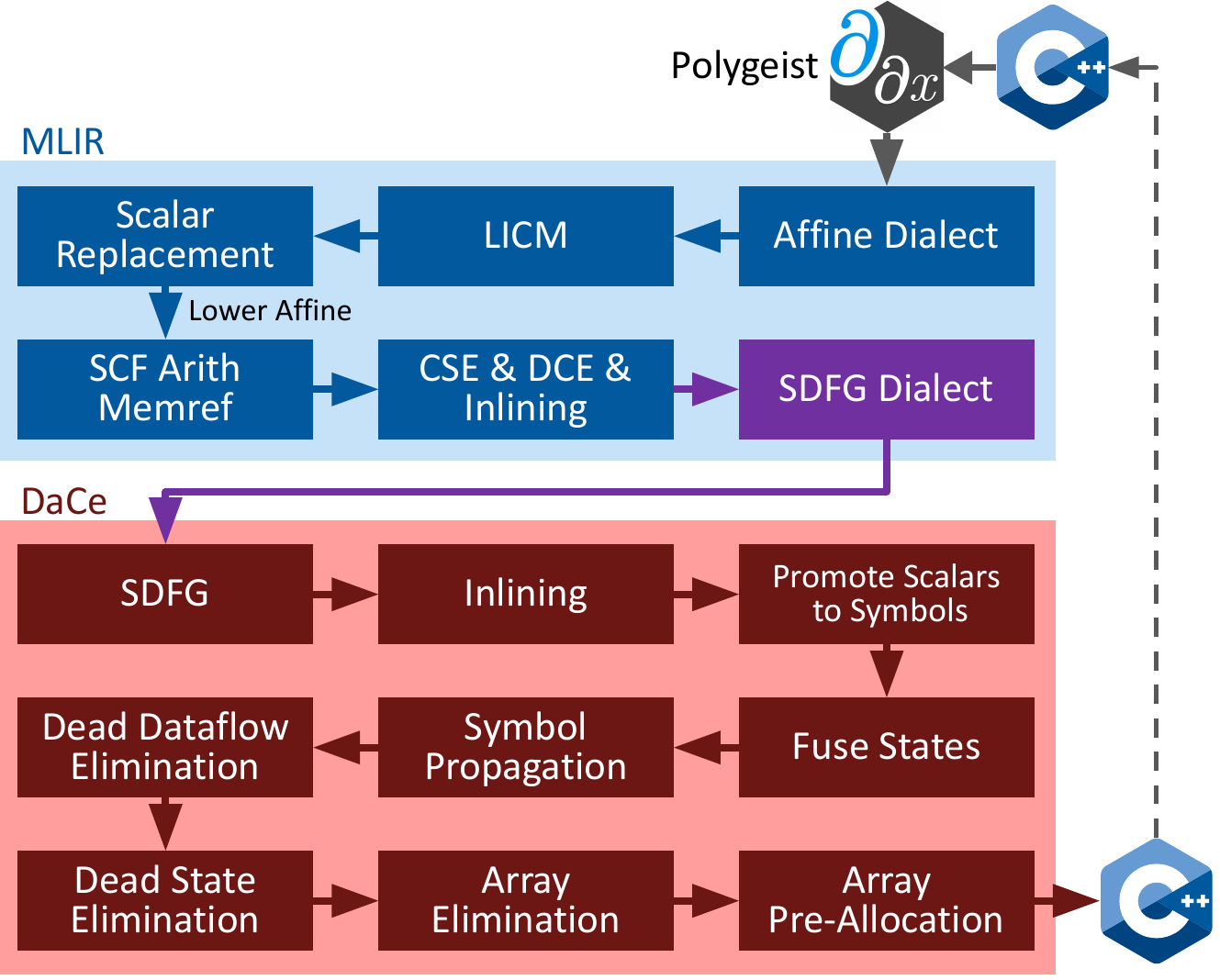}
    \vspace{-0.75em}
    \caption{Overview of the DCIR conversion pipeline.}
    \vspace{-1em}
    \label{fig:mlir-dace-overview}
\end{figure}

Due to the fundamental differences between the representations, the translation process is not direct, requiring analysis and transformation passes in the process. Fig.~\ref{fig:mlir-dace-overview} broadly describes the conversion pipeline we use, highlighting passes in the MLIR domain in blue and DaCe in red.
As DaCe generates C code, we can take the resulting representation and feed it back to MLIR for bidirectional translation.

\begin{figure*}[t]
    \centering
    \begin{subfigure}[b]{0.13\textwidth}
\begin{lstlisting}[style=tinyC]

int fName(
  int *A, 
  int *B) {

  return *A + *B;

}
\end{lstlisting}
        \vspace{-0.5em}
        \caption{Source}
        \label{fig:dialect_pipeline_c}
    \end{subfigure}
    \quad
    \begin{subfigure}[b]{0.32\textwidth}
\begin{lstlisting}[style=tinyMLIR]
module {
 func.func @fName(
   %arg0: memref<?xi32>, $\circledColorSmall{1}{black}$ 
   %arg1: memref<?xi32>) -> i32 {

  %c0 = arith.constant 0 : index
  %0 = memref.load %arg0[%c0] : memref<?xi32>
  %1 = memref.load %arg1[%c0] : memref<?xi32>
  %2 = arith.addi %0, %1 : i32
  return %2 : i32

 }
}

\end{lstlisting}
        \vspace{-0.5em}
        \caption{Polygeist-generated MLIR}
        \label{fig:dialect_pipeline_mlir}
    \end{subfigure}
    \quad
    \begin{subfigure}[b]{0.31\textwidth}
\begin{lstlisting}[style=tinyMLIR]
module {
 sdfg.sdfg (
   %arg0: !sdfg.array<sym("s_0")xi32>, $\circledColorSmall{1}{black}$
   %arg1: !sdfg.array<sym("s_1")xi32>) -> (
     %arg2: !sdfg.array<i32>) { [...]
  sdfg.state @addi_9 { [...] $\circledColorSmall{3}{black}$
   %6 = sdfg.tasklet (
     %4 as %arg3: i32, $\circledColorSmall{2}{black}$
     %5 as %arg4: i32) -> (i32) {
    %8 = arith.addi %arg3, %arg4 : i32
    sdfg.return %8 : i32
   } [...]
  } [...]
 }
}

\end{lstlisting}
        \vspace{-0.5em}
        \caption{SDFG dialect}
        \label{fig:dialect_pipeline_sdfg_dialect}
    \end{subfigure}
    \hfill
    \begin{subfigure}[b]{0.17\textwidth}
        \centering
        \includegraphics[height=1.7in,page=1]{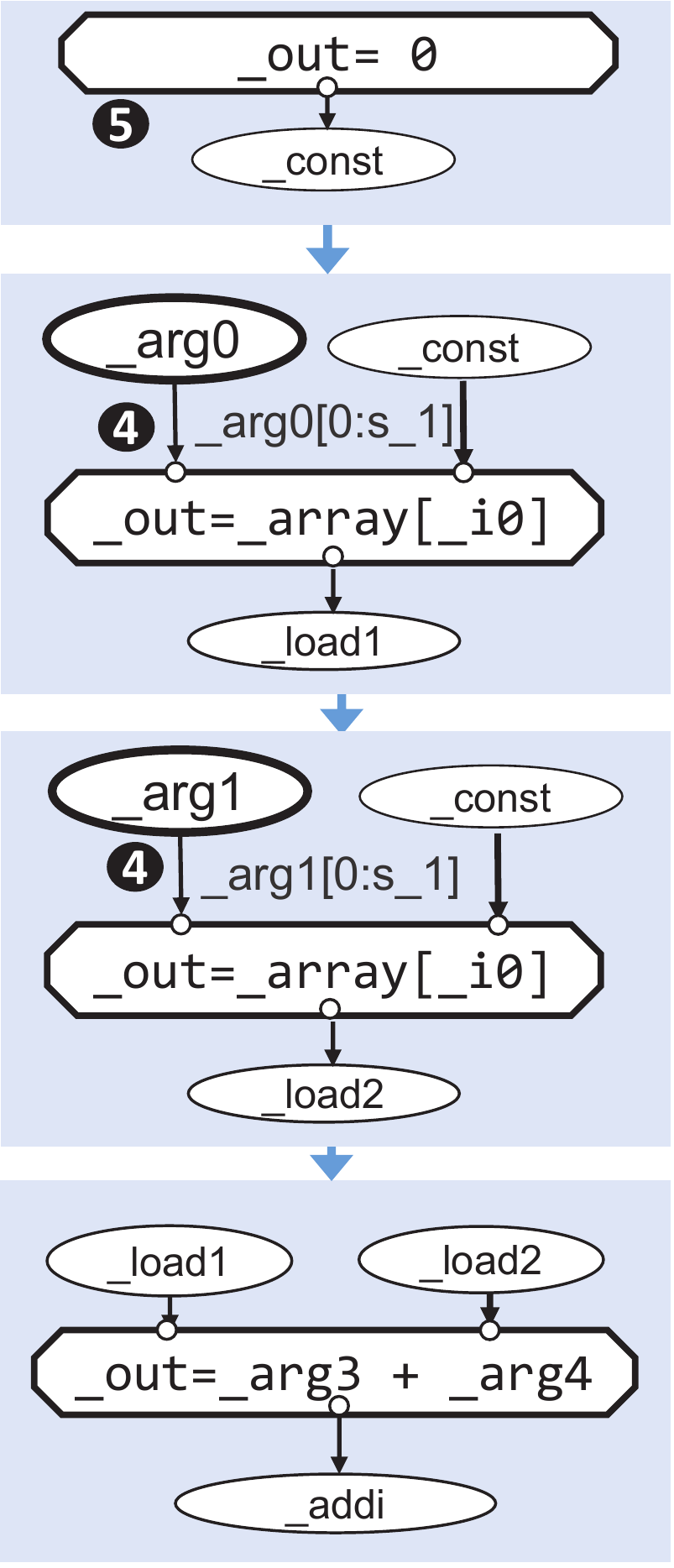}
        \vspace{-0.5em}
        \caption{Translated SDFG}
        \label{fig:dialect_pipeline_sdfg}
    \end{subfigure}
    \vspace{-1em}
    \caption{Conversion and translation of a simple program.}
    \label{fig:dialect_pipeline}
\end{figure*}

The process (demonstrated in Fig.~\ref{fig:dialect_pipeline}) starts with input code and a frontend, for which we use C/C++ and Polygeist~\cite{polygeistPACT}, which is the state of the art MLIR frontend for those languages. Our proposed pipeline starts with Polygeist generated code in the \texttt{builtin}, \texttt{func}, \texttt{memref}, \texttt{arith}, \texttt{math} and \texttt{affine} dialects. We then apply a suite of typical control-centric passes --- loop-invariant code motion, dead code elimination, common subexpression elimination, function inlining --- and lower code to adhere to the dialects \texttt{sdfg} can be converted from (Fig.~\ref{fig:dialect_pipeline_mlir}). 
The pipeline then generates the corresponding code in the \texttt{sdfg} dialect (Fig.~\ref{fig:dialect_pipeline_sdfg_dialect}), described in Section~\ref{sec:mlirconv}, and translates it to the SDFG IR (Fig.~\ref{fig:dialect_pipeline_sdfg}). Following translation, we must perform additional passes on the SDFG (\S\ref{sec:passes}) to ensure that data-centric optimizations can be applied. Lastly, we apply a suite of typical data-centric passes for data movement reduction and memory scheduling.

\section{MLIR Conversion and Translation} \label{sec:mlirconv}

To work, our figurative bridge needs to implement two interfaces on the MLIR and DaCe endpoints.
In MLIR nomenclature, these components are \textit{converters}, which transpile one dialect to another, and \textit{translators}, which interface MLIR with other representations. We now describe the design decisions and considerations made in implementing both components.

\subsection{Converter} \label{sec:converter}

A converter needs source dialects to convert from.
We choose four core source MLIR dialects --- \texttt{scf}, \texttt{arith}, \texttt{math}, and \texttt{memref} --- in order to maximize compatibility (as most frontends and external dialects specify lowering to those). The conversion process is nearly direct: (a) memory allocation and load/store operations from the \texttt{memref} dialect are converted to \texttt{sdfg.\{alloc,load,store\}}; (b) arithmetic and mathematical computations, as well as unknown operations, are converted into separate \texttt{sdfg.tasklet}s (see below); and (c) \texttt{scf} constructs are lowered to state machine subgraphs.

Any encountered question mark or star in \texttt{memref} sizes is replaced with a \textit{unique identifier}, preserving original MLIR semantics. In the example in Fig.~\ref{fig:dialect_pipeline} this can be seen with  \texttt{sym("s\_0")} (\circledColorSmall{1}{black}).  As conversion progresses on the MLIR side, the number of symbols is gradually reduced by propagating their values forward through references. The process is limited to assignments (rather than arbitrary subsets), but the process continues with DaCe's symbolic math engine as a data-centric transformation (\S\ref{sec:symbolprop}).

It is crucial to track provenance of memory operations. Whenever a load or store is encountered, it is converted to data-centric scoping --- propagating data dependencies outwards. Namely, an operation is injected in every scope (tasklet, control-flow construct, function, or parallel map/consume) it is used, and all parent scopes must specify a \textit{symbolic} subset of the data moved from the outer scope (\circledColorSmall{2}{black}). If analysis cannot determine the subregion, it is set to be equal to the outer region and refined later with DaCe's symbolic math engine.

To express computations in the \texttt{sdfg} dialect, the converter generates SDFG states and tasklets.
In order to retain program order semantics, we first place every computation in its own \texttt{sdfg.state} (\circledColorSmall{3}{black}), which may be subsequently fused in DaCe (\S\ref{sec:passes}). We also split each computational operator into an individual tasklet, allowing DaCe to recover dataflow.

Conversion in the other direction poses fewer challenges. Similarly to SDFG code generation~\cite{dace}, tasklet contents can be inlined to MLIR core dialects, structured control flow can be raised from the state machine (e.g., using dominator analysis), and symbols become scalar identifiers.

\subsection{MLIR-to-SDFG Translator}
Once the input program is in the \texttt{sdfg} dialect, it is translated to the SDFG format in two passes. The first pass collects symbol, container, and scope metadata for constructing the graph; and the second pass creates and connects the graph.

MLIR operators that are not supported by our converter are kept as-is and compiled as \textit{MLIR tasklets}. We add this functionality to DaCe by creating shims that convert data containers to \texttt{memref}s, compiling the tasklet contents with \texttt{mlir-opt} and \texttt{llc}, and linking the resulting module.

As MLIR tasklets create multiple object files, they are only optimized via link-time optimization (LTO) and may thus yield lower performance than with single translation unit passes. During translation, we resolve this by \textit{raising} MLIR tasklets to Python tasklets (which are native to DaCe) if possible. This includes parsing arithmetic operators into Python equivalents (e.g., \texttt{arith.addi \%a, \%b} to \texttt{a + b}) and built-in math library calls. Raising not only inlines the tasklet code during compilation in DaCe, but also enables data-centric analyses and transformations, including arithmetic intensity estimation and symbolic auto-differentiation~\cite{daceml}.

%% file: sections/transformations.tex
\section{Data-Centric Passes}
\label{sec:passes}

Trivial translation of MLIR to SDFG may yield a data-centric representation that is functional, but not optimizable. We therefore make additional adaptations on the resulting IR to increase data movement analysis capabilities. Subsequently, we apply automatic optimizations (\texttt{-O1} and \texttt{-O2} compiler-flag equivalent) that reduce data movement to increase application performance.

\subsection{Inference}

We begin with global inference of symbolic expressions, update operations, and increasing the size of pure dataflow regions (states). The resulting recovered information enriches dataflow analysis and enables more optimizations.

\macsection{Scalar to Symbol Promotion}
This first pass elevates scalar values into symbolic expressions, if they can be represented as such, are used in indices/shapes, and do not change during their lifetime (\circledColorSmall{4}{black}). In our example, \texttt{\_const} is elevated to symbol, and therefore  the memlet \texttt{\_arg0[0:s1]} becomes \texttt{\_arg0[\_const]}.
This is crucial for MLIR codes, as every SSA value becomes a scalar data container. With this pass, index expressions, loop bounds, data-dependent memory allocations, and other elements are readily exposed to DaCe.

\macsection{Symbolic Inference} \label{sec:symbolprop}
Given DCIR's tendency to create many symbols (e.g., for each question mark), reducing their number is crucial for validation and optimization. The \textit{symbol propagation} pass works similarly to constant propagation, forwarding values of symbolic expressions and replacing symbols if they are set once (\circledColorSmall{5}{black}). Propagation works in two passes over the code --- one to collect symbols that are assigned once (via forward and reverse reachability), and another to make the necessary replacements. In the example from Fig.~\ref{fig:dialect_pipeline_sdfg}, \texttt{\_const} is detected to be 0, and that value is then propagated so the memlet \texttt{\_arg0[\_const]} becomes \texttt{\_arg0[0]}.
Additionally, on every function call, an attempt is made to reduce symbols by solving a system of equations.

\macsection{Update Detection}
SDFGs support a third mode of data movement: update. Differentiating between updates and writes is important for several optimizations, including automatic parallelization~\cite{c2dace}, detecting and improving reduction schedules, and choosing wait-free operations such as one-sided communication. We invoke the \texttt{AugAssignToWCR} DaCe transformation to detect updates via symbolic expression tracing around tasklets. If a read and a write operate on the same memory location, and it is expressible by an associative binary function, the read/write is converted to an update.

\macsection{SDFG Simplification}
The simplification pipeline~\cite{dace} is an idempotent process that repeatedly fuses control-flow elements (states and nested CFGs) to enlarge pure dataflow regions. This functionality is given via the built-in DaCe API, which we invoke using \texttt{sdfg.simplify()}. More specifically, the method fuses SDFG states if their data dependencies can be expressed in one acyclic graph without introducing data races; and hoists control-flow that is situated inside dataflow (such as a branch inside parallel dataflow sections).

\vspace{0.25em}Once symbolic information is appropriately annotated, we can reduce unnecessary copies and memory allocation.

\subsection{Data Movement Reduction (\texttt{-O1})}

In this work, we introduce novel data-centric passes into the DaCe framework, in order to support conversion of arbitrary MLIR codes to efficient programs.

\macsection{Extended Dead Code Elimination (DCE)} As a bridging pass between control- and data-centric transformations, we extend the notion of DCE in DaCe. Global dataflow and symbolic analysis lend their usefulness in DCE, which we can use to eliminate complete sequences of array operations. We perform the novel DCE in two separate passes: Dead State Elimination and Dead Dataflow Elimination. The former uses the propagated symbols to determine whether an expression will always be false, and eliminates unreachable state machine states. Dead Dataflow Elimination then traverses the state machine in reverse topological order, tracking future-reused data containers (arrays, scalars) and removing all computations that end up in unused temporary containers.

\macsection{Array Elimination}
``Dead memory'' does not only refer to unused arrays, but also to extraneous memory copies and subregions that remain unused. DaCe can already detect redundant copies~\cite{dace}, but we extend this feature from a local transformation to a pass that reduces memory usage via removing arrays and views through a linear-time traversal.

\macsection{Memlet Consolidation} After converting MLIR dialects and propagating data dependencies (\S\ref{sec:converter}), we may end up with multiple memlets that refer to overlapping regions. For example, a stencil that reads \texttt{A[i]} and \texttt{A[i + 1]} would generate two separate memlets. 
We thus define a pass that unions memlets that refer to the same containers within the same scope, as a form of data movement common denominator.

\subsection{Memory Scheduling Optimizations (\texttt{-O2})}

Our final step in the pipeline is to optimize the program and its order based on its relationship with its memory. In particular, we apply two optimizations on the program schedule:

\macsection{Memory (Pre-)Allocation}
DaCe already controls memory allocation and deallocation implicitly, based on data containers' lifetime. However, it can still yield suboptimal performance with arbitrary MLIR codes, e.g., if an array is allocated within a critical loop, causing many unnecessary calls. Two passes deal with this: one heuristic analyzes the program, deciding if a container could be allocated on the stack or registers rather than the heap; the other heuristic moves memory allocation to the outermost scope it can (if no data races occur), removing such calls from the critical path.

\macsection{Memory-Reducing Loop Fusion}
The inferred explicit, reduced symbolic expressions allow us to analyze computational schedules and mutate them to further minimize data movement. We greedily invoke the existing fusion transformations available in DaCe as a heuristic pass. Those transformations merge scopes that write and read from otherwise-unused intermediate data, if memory access pattern permits to do so. This reduces the size of the intermediate array to a scalar (or the common subregion, e.g., if sparse), promoting cache locality and reducing memory footprint.

\vspace{0.5em}
Overall, the set of optimizations we perform in the pipeline is conservative.
Other scheduling and footprint reduction passes (e.g., relayouting, memory pooling) can be beneficial for allocation but harm cache locality, thus they are only enabled manually.
As we shall show, the above set still yields promising performance gains on general applications.

%% file: sections/evaluation.tex
\begin{figure*}[t]
    \centering
    \includegraphics[width=.95\textwidth]{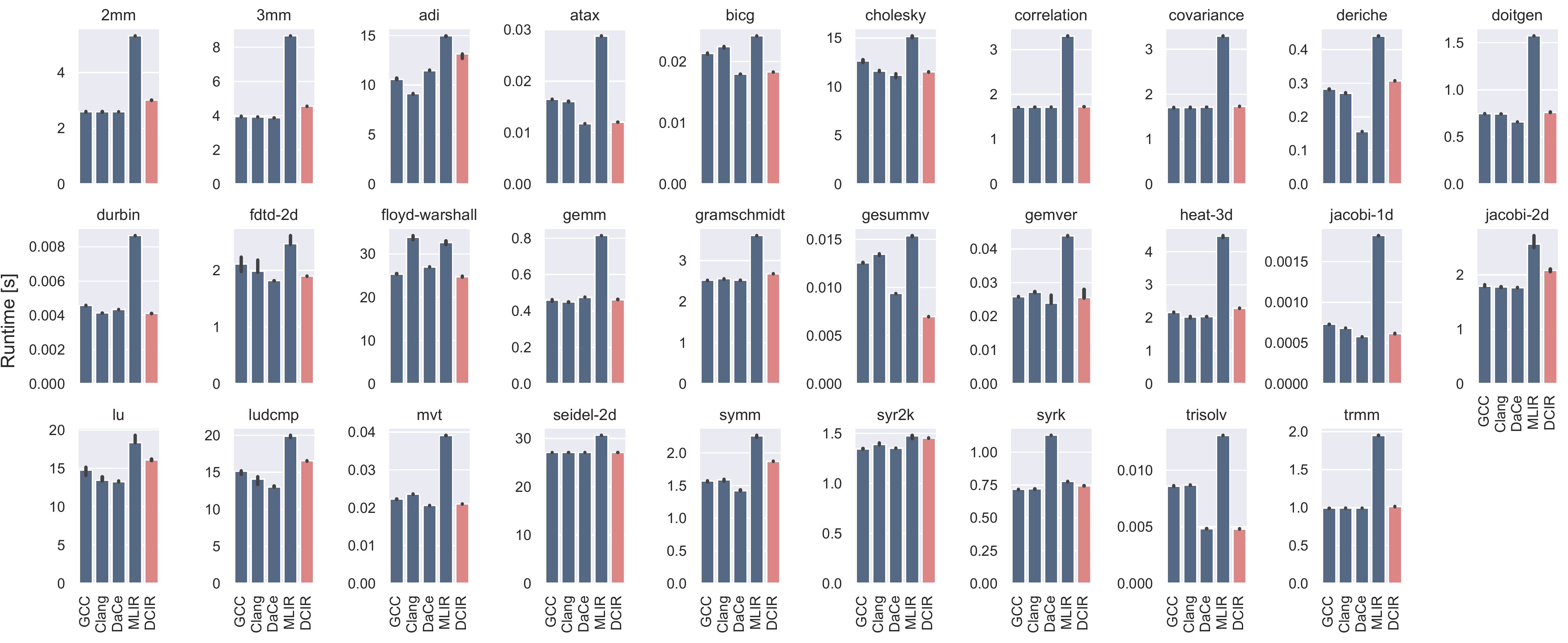}
    \vspace{-1em}
    \caption{Polybench/C Benchmark comparing GCC, Clang, the DaCe C frontend, and MLIR (via Polygeist) with DCIR. Geometric mean of DCIR speedups: 1.59$\times$ over Polygeist with MLIR, 1.03$\times$ over GCC, 1.02$\times$ over Clang, 0.94$\times$ over DaCe.}
    \label{fig:evaluation_benchmarks}
\end{figure*}

\section{Evaluation}  \label{sec:evaluation}
In this section, we show that a single pass through our pipeline can generate code that matches or outperforms established compilers. 

\subsection{Experimental Setup}  \label{sec:setup}
For our evaluation, we use the DCIR pipeline, described in Section~\ref{pipeline}, without feeding the generated code back into the pipeline. We use DaCe (version 0.14), and disable auto-parallelization in order to ensure a uniform execution environment for all compilers and prevent conflating memory benefits with parallel efficiency. Unless otherwise mentioned, we use Clang++\footnoteref{llvm-commit} to compile the generated C++ code from DaCe. We run all of our benchmarks on a server that contains an Intel 16-core Xeon Gold 6130 CPU (clocked at 2.80~GHz) and 1.5~TB DDR4 RAM. The server consists of 32~KB of L1 caches, 1~MB of L2 caches, and a 22~MB L3 cache.

We compare the generated code with the established compilers GCC (version 12.1.0) and Clang\footnote{\label{llvm-commit}Commit: 00a12585933ef63ff1204bf5cd265f0071d04642} with the \texttt{-O3 -fPIC -march=native} flags. For one application (\texttt{gramschmidt}), we instead use the \texttt{-O2} flag, as due to the numerical sensitivity of the application, the baseline compilers generate wrong results on \texttt{-O3}. 
Additionally, we run the benchmarks directly using the DaCe C frontend\footnote{\label{c2dacecommit}Commit: e7fe56262ba20b6e7e664eef169dcf32b1907be9} using the same flags. Performance counters were measured with PAPI~\cite{papi} 7.0.0.

We generate MLIR\footnoteref{llvm-commit} code from C via Polygeist\footnote{Commit: fc15676b30c80ac1adb10f4fc3e7d7e8fe3ed7b6}, applying the same optimization passes as in our pipeline and lowering it to the LLVM IR, which we compile using \texttt{llc}\footnoteref{llvm-commit} and Clang\footnoteref{llvm-commit}, with the same flags. This not only provides another pipeline to compare with, but also allows separating the performance gains of the control-centric and data-centric optimizations.

\subsection{Fundamental Computational Kernels} \label{sec:polybench}
In order to evaluate our pipeline on practical applications, we choose the Polybench/C~\cite{polybench} benchmark (version 4.2.1) because it contains numerical computations with static control flow from various application domains, such as linear algebra, image processing, physics simulation, dynamic programming, and statistics. This provides a wide range of practical applications and enables a more general evaluation. Since Polybench kernels allocate all memory in advance, it is expected that the performance without schedule changes would be at the same level as the best compiler.

We use the benchmarks in their unaltered form, apart from increasing the printing precision to more accurately check the correctness of the outputs.  
We disable the auto-optimization pass for \texttt{doitgen}, \texttt{durbin}, and \texttt{gemver} because the pass produced wrong results due to its experimental status. For \texttt{floyd-warshall} we had to remove both the simplification and auto-optimization passes as they both raised exceptions. Instead, we apply a subset of those passes: optional array inference, symbol propagation, state fusion, and memory (pre-)allocation.
Furthermore, we exclude \texttt{nussinov} from the benchmarks, as Polygeist was unable to generate MLIR core dialect operations for the entire application.

All benchmarks were run ten times on the large dataset defined by Polybench using double-precision floating-point numbers. The plots in Fig.~\ref{fig:evaluation_benchmarks} represent the median of the measured runtimes with 95\% confidence intervals. In our benchmarks, we measured the runtimes of the whole applications, contrary to Polybench solely measuring the execution time of the kernels. This allows us to additionally consider program-wide data-centric optimization passes.

Compilation time for each of the benchmarks ranges between 19--64~seconds (median: 24~s), where the median DCIR optimization time is 3.46~s, and most of the time (13.94~s) is spent in CMake, which DaCe calls on the generated code.

DCIR strictly outperforms the MLIR pipeline by a factor of 1.59$\times$ on average, on par with the best general-purpose compilers. We highlight four key observations:
\vspace{-.1em}
\begin{enumerate}
    \item DCIR is never slower than MLIR, showing that it retains the control-centric optimizations of MLIR.
    \item DCIR strictly outperforms the MLIR pipeline, proving that there is performance to be gained by applying data-centric optimizations. 
    \item DCIR is on par with GCC and Clang w.r.t. the geometric mean speedup. We thus successfully recover the performance that the MLIR pipeline leaves untapped. 
    \item The Polybench suite does not contain extraneous arrays or unnecessary regions, which would enable additional data-centric optimizations, so there is no significant speedup compared with GCC and Clang.
\end{enumerate} 

The cases with increased performance can be attributed to moving arrays to the stack and to improved scheduling. 
For example, on \texttt{gesummv}, DCIR moves one out of the five arrays to the stack, which improves the execution time for loading and storing operations.
These benchmarks demonstrate the possible speedup provided by the novel data-centric optimization passes.

\begin{figure}[t]
    \centering
    \begin{subfigure}[b]{0.47\linewidth}
\begin{lstlisting}[style=tinyC]
for (i = 0; i < _PB_N; i++){
 for (k = 0; k < _PB_M; k++) {
  for (j = 0; j <= i; j++)
   C[i][j] += alpha * A[i][k] 
            * A[j][k];
 }
}
\end{lstlisting}
        \vspace{-0.5em}
        \caption{Source}
        \label{fig:evaluation_syrk_orig}
    \end{subfigure}
    \hfill
    \begin{subfigure}[b]{0.47\linewidth}
\begin{lstlisting}[style=tinyC]
for (i = 0; i < _PB_N; i++){
 for (k = 0; k < _PB_M; k++) {
  tmp = alpha * A[i][k]
  for (j = 0; j <= i; j++)
   C[i][j] += tmp * A[j][k];
 }
}
\end{lstlisting}
        \vspace{-0.5em}
        \caption{DCIR-generated C++ code}
        \label{fig:evaluation_syrk_dcir}
    \end{subfigure}
    \vspace{-1em}
    \caption{\texttt{syrk} kernel on DaCe and DCIR.}
    \vspace{-1em}
    \label{fig:evaluation_syrk}
\end{figure}

Furthermore, we notice that the DaCe C frontend underperforms on the symmetric rank-k update (\texttt{syrk}) benchmark compared to all the other compilers and pipelines. In Fig.~\ref{fig:evaluation_syrk_orig} we can see a snippet of the \texttt{syrk} benchmark kernel. Because \texttt{alpha * A[i][k]} is completely independent of \texttt{j}, we can move it out of the innermost loop, which can be seen in Fig.~\ref{fig:evaluation_syrk_dcir}. The DaCe C frontend misses this optimization, because the generated tasklets are indivisible C++ codes --- and these tasklets are treated as black box units of computation, preventing internal optimizations.

The direct C-to-SDFG parser (DaCe) outperforms DCIR with a geometric mean speedup of 1.04$\times$. This reduced performance can be seen most prominently on \texttt{deriche}. We can observe that DaCe outperforms DCIR on \texttt{deriche} by a factor of 1.7$\times$, which is likely due to Polygeist and MLIR performing loop order inversion --- transforming loops iterating using decrements into loops using increments\footnote{The \texttt{scf} dialect is \textit{inherently limited} as it defines the step of a for-loop as a strictly positive integer.}. Indeed, performance counters indicate that DCIR exhibits 2.41$\times$ L2 and 2.72$\times$ L3 cache misses over the direct C parser. 
This benchmark exemplifies the semantic information lost in the Polygeist-MLIR pipeline.

In general, the combined DCIR pipeline outperforms the MLIR standard \texttt{-O3} mode, resulting in it being on par with Clang and GCC, \textit{all while retaining the flexibility to work with arbitrary high-level IRs}.

\subsection{Case Studies}
In addition to the Polybench benchmarks, we use code snippets from real-world applications and compare their performance across the same compilers. We only alter the code snippets in order to run them in an isolated environment. 

\begin{figure}[t]
    \centering
    \begin{subfigure}{.8\linewidth}
\begin{lstlisting}[style=tinyPy]
class Mish(nn.Module):
    def __init__(self):
        super().__init__()

    def forward(self, x):
        x = torch.log(1 + torch.exp(x))
        return x
\end{lstlisting}
        \vspace{-0.5em}
        \caption{Input Python code}
        \label{fig:evaluation_rwa_1_code}
    \end{subfigure}
    \begin{subfigure}{\linewidth}
        \centering
        \includegraphics[width=.8\textwidth]{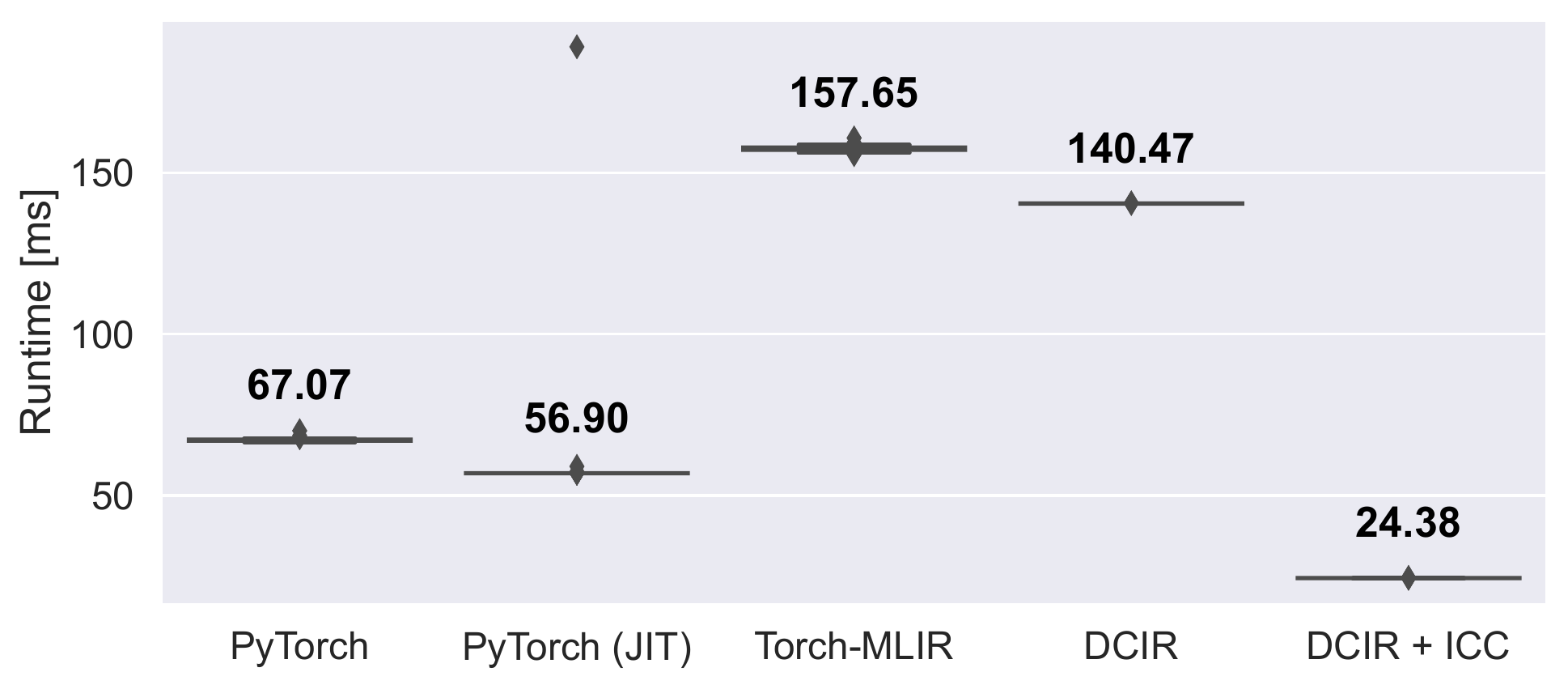}
        \caption{Performance across compilers}
        \label{fig:evaluation_rwa_1_results}
    \end{subfigure}
    \vspace{-1em}
    \caption{The Mish operator in PyTorch and its performance.}
    \vspace{-1em}
    \label{fig:evaluation_rwa_1}
\end{figure}

Fig.~\ref{fig:evaluation_rwa_1_code} shows the Mish~\cite{mish} activation function, which is used in object detection deep neural networks~\cite{yolov4}, written in PyTorch. The LLVM Torch-MLIR\footnote{\url{https://github.com/llvm/torch-mlir}} project takes deep neural networks written in PyTorch and compiles them through MLIR's core dialects using the linalg-on-tensors and MHLO (also used in TensorFlow) dialects. 

We benchmark PyTorch 1.14, its built-in \texttt{torch.jit} compiler, Torch-MLIR's optimizing pipeline, and DCIR, plotting the results in Fig.~\ref{fig:evaluation_rwa_1_results}. PyTorch, being an eager-execution framework, uses separate tensors for each intermediate value. The \texttt{torch.jit}-compiled version reports that it fused operators, mitigating the call overhead and reducing intermediate tensor allocation. We find that Torch-MLIR's generated IR also contains allocation operations, which add extraneous data movement and inhibit rescheduling. Running the DCIR pipeline removes \textit{all} allocation calls and is able to successfully fuse the computations, improving the performance by 12\% over the standard pipeline. We additionally noticed that Clang does not vectorize math library calls (namely, \texttt{exp} and \texttt{log}). Since PyTorch internally uses the SLEEF vector math library~\cite{sleef}, we also compile the DCIR-generated code with the Intel C Compiler (ICC 2021.3). The produced binary contains vector math operations of the same length, and, combined with data movement reduction, outperforms the state of the art (JIT-compiled PyTorch) by 2.33$\times$.

\begin{figure}[t]
    \centering
    \begin{subfigure}{.81\linewidth}
\begin{lstlisting}[style=tinyC]
[...]
  for (j = 1; j < Norder; j++) {
    if (converged[j] == 0) {
      zeta_ip1[j] = zeta_i[j] * zeta_im1[j] 
                  * beta_im1[0];
      c1 = beta_i[0] * alpha[0] * (zeta_im1[j] 
            - zeta_i[j]);
      c2 = zeta_im1[j] * beta_im1[0] * (1.0 - (
          shift[j] - shift[0]) * beta_i[0]);
      zeta_ip1[j] /= c1 + c2;

      beta_i[j] = beta_i[0] * zeta_ip1[j] / zeta_i[j];
    }
  }
[...]
\end{lstlisting}
        \vspace{-0.5em}
        \caption{Input code}
        \label{fig:evaluation_rwa_2_code}
    \end{subfigure}
    \begin{subfigure}{\linewidth}
        \centering
        \includegraphics[width=.8\textwidth]{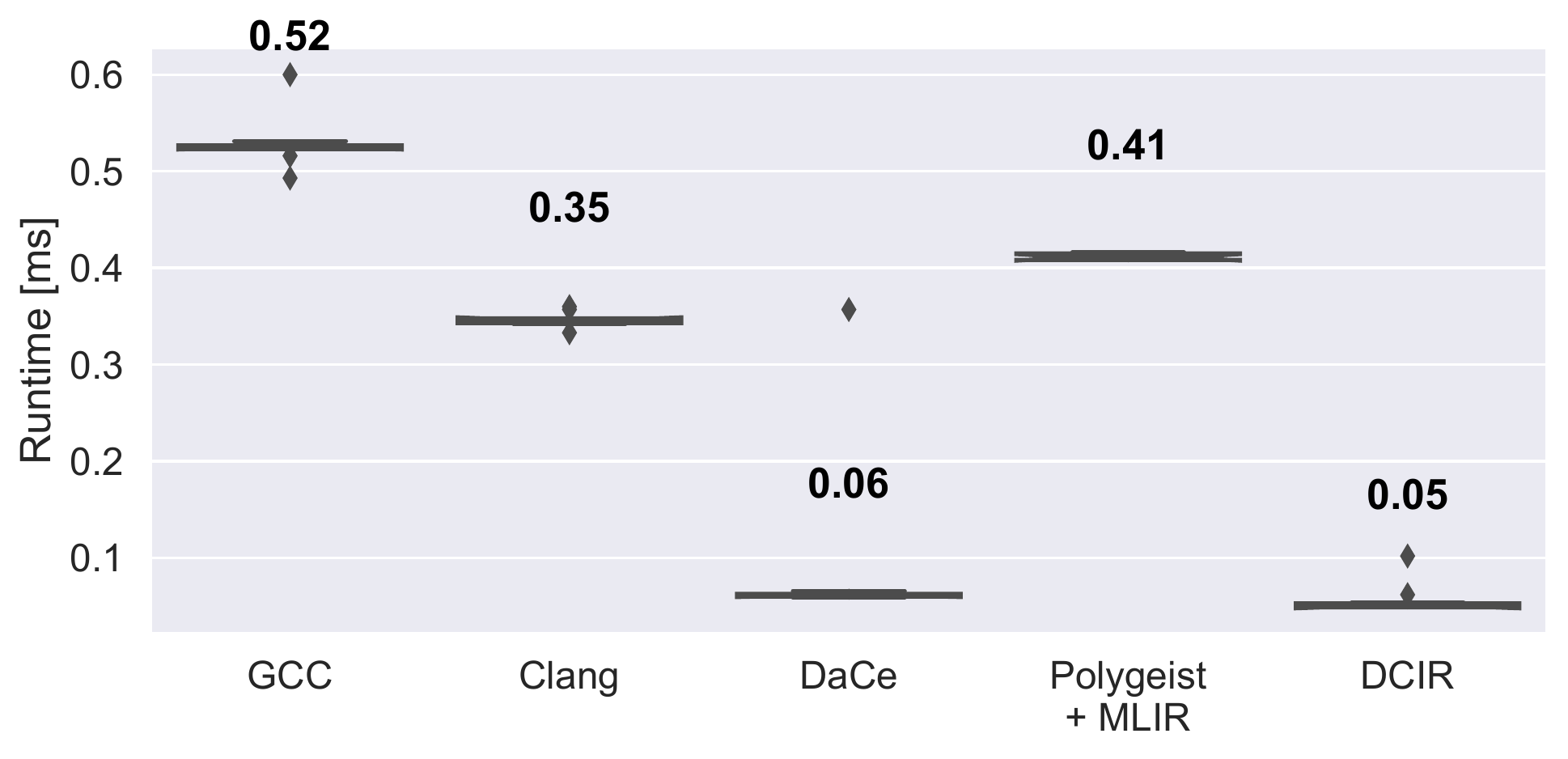}
        \caption{Performance across compilers (speedups: 8.4$\times$ over Polygeist with MLIR, 10.4$\times$ over GCC, 7$\times$ over Clang, 1.2$\times$ over DaCe)}
        \label{fig:evaluation_rwa_2_results}
    \end{subfigure}
    \vspace{-1em}
    \caption{MILC codebase case study performance.}
    \vspace{-1em}
    \label{fig:evaluation_rwa_2}
\end{figure}

For our second case study, we use one of the solvers in the MILC lattice quantum chromodynamics scientific application\footnote{\url{https://github.com/milc-qcd/milc_qcd/blob/master/arb_overlap/congrad_multi_field.c}}. The snippet in Fig.~\ref{fig:evaluation_rwa_2_code} is an implementation of a multi-mass conjugate gradient algorithm, which is an iterative method for solving sparse systems of linear equations.

Fig.~\ref{fig:evaluation_rwa_2_results} shows that DCIR achieves a speedup of 7$\times$ compared with general-purpose compilers. The snippet contains multiple array allocations and primarily consists of computing and moving their entries. DCIR and DaCe apply data-centric optimizations, eliminating two arrays, each containing 10,000 doubles, which explains the performance increase.

\begin{figure}[t]
    \centering
    \begin{subfigure}{.8\linewidth}
\begin{lstlisting}[style=tinyC]
[...]
 for (int k = 0; k < NTIMES; k++) {
  for (int i = 0; i < N; i++)
   a[i] = scalar;
  tmp = a[10];
  double sum = 0.0;
  for (int i = 0; i < N; i++)
   sum += a[i];
  a[10] = sum;
  a[10] = tmp;
  for (int i = 0; i < N; i++)
   a[i] = a[i] * scalar;
 }
[...]
\end{lstlisting}
        \vspace{-0.5em}
        \caption{Input code}
        \label{fig:evaluation_rwa_3_code}
    \end{subfigure}
    \begin{subfigure}{\linewidth}
        \centering
        \includegraphics[width=.8\textwidth]{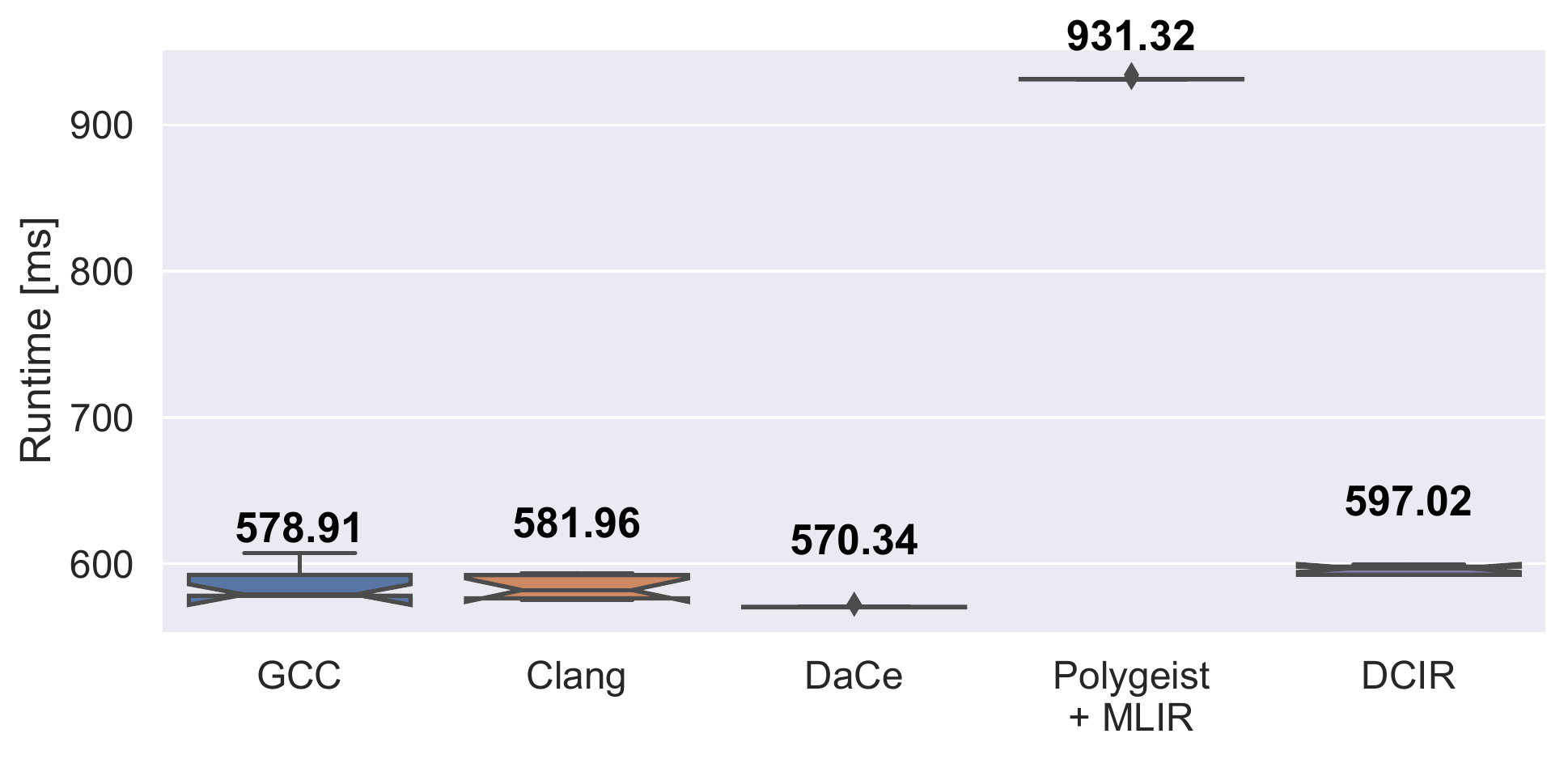}
        \caption{Performance across compilers (speedups: 1.56$\times$ over Polygeist with MLIR, 0.97$\times$ over GCC and Clang, 0.96$\times$ over DaCe)}
        \label{fig:evaluation_rwa_3_results}
    \end{subfigure}
    \vspace{-1em}
    \caption{Memory bandwidth case study performance.}
    \vspace{-1em}
    \label{fig:evaluation_rwa_3}
\end{figure}

For our third benchmark, we use a memory bandwidth benchmarking repository\footnote{\url{https://github.com/RRZE-HPC/TheBandwidthBenchmark}}, which we can see in Fig.~\ref{fig:evaluation_rwa_3_code}. This benchmark consists of allocating four arrays and performing computations, such as initializing the arrays,
multiplying all entries with a scalar or summing all elements.

We can see in Fig.~\ref{fig:evaluation_rwa_3_results} that DCIR achieves a speedup of 1.56$\times$ compared with MLIR, and is on-par with GCC and Clang.
As the results indicate, DCIR again recovers the performance lost by the MLIR-generated code.

In total, 63 arrays and scalars were eliminated from the three snippets, contributing most to the observed speedup.

In conclusion, both C applications in the wild and neural networks benefit from data-centric optimizations. DCIR not only successfully applies those, but also recovers semantics lost during conversion of the source language to MLIR, regaining or surpassing the performance of general compilers.

%% file: sections/related_work.tex
\section{Related Work}

In this section, we discuss other approaches that share some of the same goals or methods as our own.

\macsection{Compilers with multiple abstraction layers} In this work, we focus on the Polygeist and Torch-MLIR frontends, but other frontends are being developed, which will likely become further candidates for our approach. Notable examples of emerging MLIR frontends are Flang\footnote{\url{https://github.com/flang-compiler/flang}} and
Pylir\footnote{\label{pylir}\url{https://github.com/zero9178/Pylir}}. At the moment of writing, these frontends use MLIR internally but lack the capability of emitting MLIR core dialect code directly. Once this changes, they could be used as a starting point to our workflow, expanding the scope of programs that can take advantage of DCIR.

SYCL~\cite{sycl} is a C++ cross-platform abstraction layer offering performance portability by leveraging template programming and lambda functions to create specialized code for different hardware architectures from the same high-level code. While not as extensible as MLIR, SYCL offers a way for applications to reach high performance on heterogeneous hardware without needing extensive rewrites. To bridge the two, SYCLops~\cite{syclops} converts SYCL LLVM IR into the \texttt{affine}, \texttt{arith}, and \texttt{memref} MLIR dialects.

\macsection{Dataflow Representations}
Several other data-centric representations exist, such as PDG~\cite{pdg}, HPVM~\cite{hpvm}, Bamboo~\cite{bamboo}, Dryad~\cite{dryad}, and Naiad~\cite{naiad}. SDFGs are unique because they encapsulate fine-grained data dependencies and differentiate between reads, writes and updates on memory. This distinction enables certain transformations which rely on such accesses. A further differentiating feature is the symbolic representation and tracking of memory access patterns throughout the application, a critical aspect that is key to understanding and optimizing dataflow.

Possibly the closest alternative to DCIR is the DaCe C frontend~\cite{c2dace}, a tool that generates SDFGs from C code. While it implements a number of AST transformations to pre-process the input program and make the translation to SDFG possible, it does not perform control-centric optimizations, nor enjoys the community support that the MLIR and LLVM environments provide. Therefore, it is unlikely that the DaCe C frontend could keep up with the versatile and domain-specific optimizations MLIR and its dialects provide.

%% file: sections/conclusion.tex
\section{Conclusion}
The paper shows how data movement minimization can aid in general-purpose program optimization. Through the use of an MLIR dialect as a bridge between the SDFG IR and MLIR core dialects, data-centric representations become commutable with control-centric IRs on any source language that MLIR supports.
The resulting performance on benchmarks and snippets from real codebases demonstrate that there is a necessity in optimizations such as dead memory elimination, and that the performance benefits can be substantial, at times exhibiting orders of magnitude of improvement.

%% file: sections/acknowledgements.tex
\begin{acks}
This project received funding from the European Research Council \includegraphics[height=.7em]{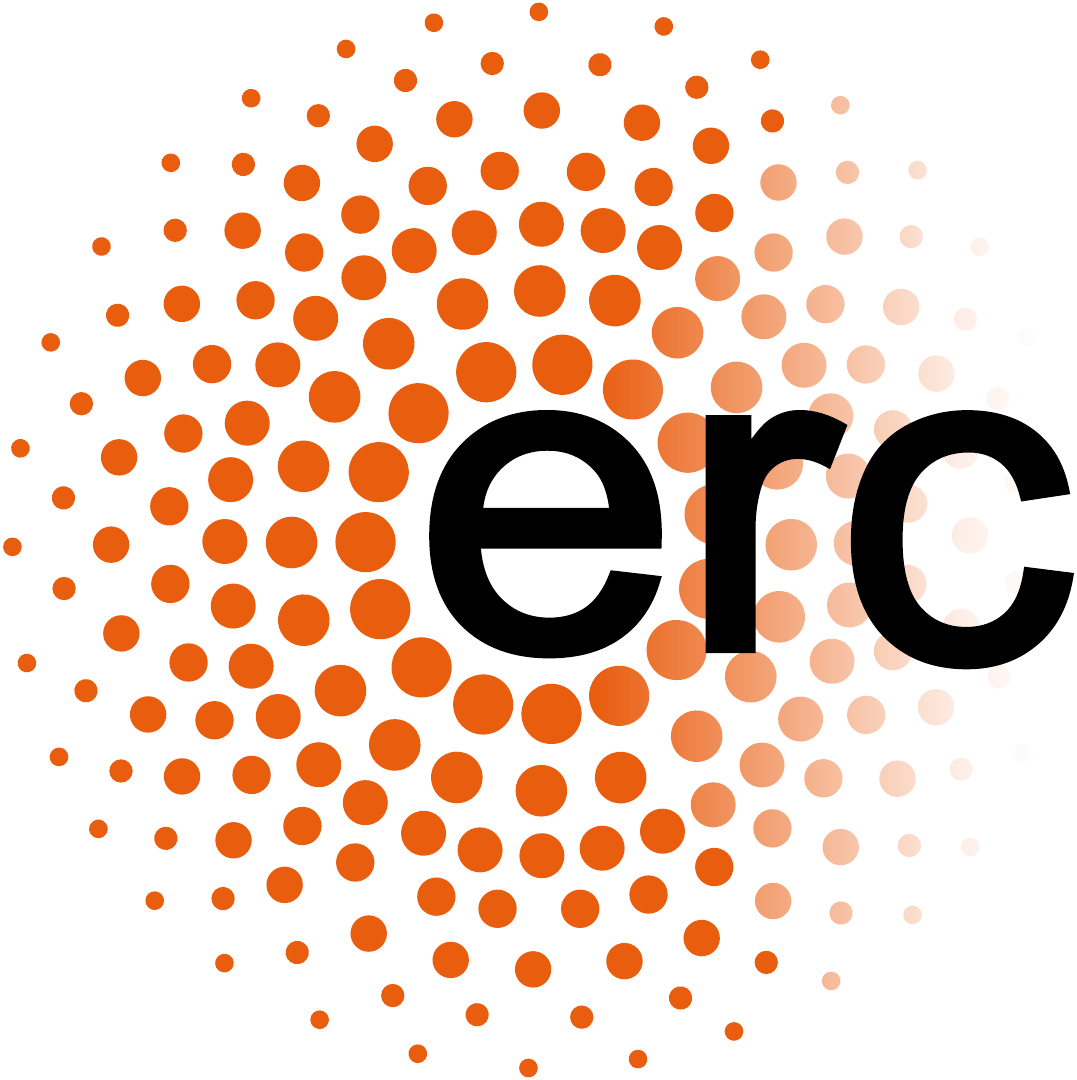} under the European Union's Horizon
2020 programme (Project PSAP, No. 101002047); and receives funding from EuroHPC-JU under grant DEEP-SEA, No. 955606, with support from the Horizon 2020 programme.
T.B.N. is supported by the Swiss National Science Foundation (Ambizione Project \#185778).
The authors also wish to acknowledge the support from the PASC program (Platform for Advanced Scientific Computing) for the DaCeMI project. This work was partially supported by the ETH Future Computing Laboratory (EFCL), financed by a donation from Huawei Technologies. 
We acknowledge the Swiss National
Supercomputing Centre (CSCS) for access to the computational resources.
\end{acks}

%% file: sections/artifact.tex
%
%
%
%
%






\appendix
\section{Artifact Appendix}

\subsection{Abstract}

The artifacts of this paper reproduce the results from Fig.~\ref{fig:introduction_motivation} and the benchmarks in Section~\ref{sec:evaluation}. The results include running the programs through the proposed compilation pipeline and its competitors, as well as outputting raw results and regenerating the figures in the paper. The benchmarks should run on any regular hardware, and a Docker container is provided for increased reproducibility.
\emph{All used software is publicly available and included in the container}.

\subsection{Artifact Check-list (Meta-information)}


{\small
\begin{itemize}
  \item {\bf Program:} PolyBench/C (Ver. 4.2.1 beta)
  \item {\bf Compilation:} GCC (Ver. 11.3.0), Clang (Ver. 13.0.1), ICC (Ver. 2021.7.1), Polygeist (git revision \texttt{e703db1})\footnote{\url{https://github.com/llvm/Polygeist/tree/e703db1}}, Torch-MLIR (git revision \texttt{eec9a7e})\footnote{\url{https://github.com/llvm/torch-mlir/tree/eec9a7e}}, DaCe (git revision \texttt{c20e8ce})\footnote{\url{https://github.com/Berke-Ates/dace/tree/c20e8ce}}
  \item {\bf Transformations:} MLIR Optimizer and Translator (git revision \texttt{00a1258})\footnote{\url{https://github.com/llvm/llvm-project/tree/00a1258}}, MLIR-HLO (git revision \texttt{2c4a384})\footnote{\url{https://github.com/tensorflow/mlir-hlo/tree/2c4a384}}, DCIR Optimizer and translator (git revision \texttt{91f067d})\footnote{\url{https://github.com/spcl/mlir-dace/tree/91f067d}}
  \item {\bf Run-time environment:} Tested on Ubuntu 22.04 and 22.10. Depending on Docker installation, root access may be required.
  \item {\bf Metrics:} Runtime of generated code
  \item {\bf Output:} CSV of runtimes for plots, raw paper results also included
  \item {\bf Experiments:} Multiple scripts to run the experiments are provided
  \item {\bf How much disk space required (approximately)?:}\\3.8~GiB to pull Docker container (12.4~GiB uncompressed), $\approx$84~GiB to build container
  \item {\bf How much time is needed to complete experiments (approximately)?:} Up to 5 hours for 10 repetitions and all compilers
  \item {\bf Publicly available?:} Yes
  \item {\bf Code licenses (if publicly available)?:} BSD 3-Clause License
  \item {\bf Workflow framework used?:} No
  \item {\bf Archived (provide DOI)?:}\\ \url{https://doi.org/10.5281/zenodo.7519936}
\end{itemize}
}

\subsection{Description}

\subsubsection{How to Access}

All the required files as well as instructions on how to execute them can be accessed on GitHub (\url{https://github.com/Berke-Ates/dcir-artifact}) and Zenodo~\cite{artifact}. The files themselves require 8.8~GiB of disk space, whereas building the Docker image from scratch additionally requires approximately 84~GiB of disk space (due to LLVM build files).


\subsubsection{Software Dependencies}
Running and building the Docker container requires an installation of Docker and a running instance of the Docker daemon.



\subsection{Installation}

There are three options for installation:
\begin{itemize}
    \item Pull the docker image
    \item Manually build the docker image
    \item Manual installation
\end{itemize}

All three options are described in the \texttt{README.md} of the repository. The first option is recommended and can be invoked by: \texttt{docker pull berkeates/dcir-cgo23:latest}

\subsection{Experiment Workflow}

For our DCIR benchmarks, we first generate MLIR code from the C source code using Polygeist. After applying various optimizations within MLIR, we convert the MLIR code to an SDFG. Using DaCe we apply further optimizations, generate C++ code, which we compile using Clang (or ICC for torch-mlir). We then measure the execution time of the generated executable and compare it with GCC, Clang, Polygeist + MLIR and DaCe.

All the compilers, translators, and code generators are built as part of the Docker container, which can be run via \texttt{docker run -it berkeates/dcir-cgo23}.

\subsection{Evaluation and Expected Results}

The \texttt{run\_all.sh} script inside the \texttt{scripts} folder will execute all benchmarks and generate the raw data as well as the plots seen in this paper. Calling the script with an output directory (for example, \texttt{./scripts/run\_all.sh ae 10} outputs to the \texttt{ae} folder with 10 repetitions) would result in CSV files and PDF files called \texttt{fig\#.pdf} for the appropriate figure number, or benchmark name for additional benchmarks. On similar hardware (see \S~\ref{sec:setup}), the plots are expected to approximately match the ones in the paper, which can be found in the artifact under the \texttt{output} directory.

\subsection{Experiment Customization}

Other benchmarks can be run by adding them in the same scheme as the ones already provided. The repository contains instructions on how to run a single benchmark. 





